\documentclass[final,1p,review,times]{elsarticle}
\usepackage{amssymb}
\usepackage{amsmath}
\usepackage{algorithm}
\usepackage{bm}
\usepackage{array}
\usepackage{algpseudocode}
\usepackage{stmaryrd}
\usepackage{lipsum}
\usepackage{subcaption}
\usepackage[toc,page]{appendix}
\usepackage{xcolor}
\newcommand{\revG}[1]{\textcolor{black}{#1}}
\newcommand{\revgg}[1]{\textcolor{black}{#1}}
\journal{Signal Processing}

\begin{document}

\begin{frontmatter}



\title{Enhanced channel estimation for double RIS-aided MIMO systems using coupled tensor decompositions}


\author{Gerald C. Nwalozie\corref{cor1}\fnref{label2} } %% Author name
\ead{gerald-chetachi.nwalozie@tu-ilmenau.de}
 %\fntext[label2]{G. C. Nwalozie, and M. Haardt are with the Communications Research Laboratory (CRL), TU Ilmenau, Ilmenau, Germany (e-mail: \{gerald-chetachi.nwalozie, martin.haardt\}@tu-ilmenau.de)}
 \cortext[cor1]{Corresponding author at: Communications Research Laboratory (CRL), Technical University Ilmenau, Ilmenau, Germany}

\affiliation[label2]{organization={Communications Research Laboratory (CRL), Technical University Ilmenau},%Department and Organization
            %addressline={}, 
            city={Ilmenau},
            %postcode={}, 
            %state={},
            country={Germany}}
\author{André L. F. de Almeida\fnref{label3}} %% Author name
%\ead{andre@gtel.ufc.br}
 %\fntext[label3]{A. L. F. de Almeida is with the Wireless Telecom Research Group (GTEL), Federal University of Cear\'a, Fortaleza, Brazil (e-mail: andre@gtel.ufc.br). }
 %\cortext[cor1]{}

\affiliation[label3]{organization={Wireless Telecom Research Group (GTEL), Federal University of Ceara},%Department and Organization
            %addressline={}, 
            city={Fortaleza},
            %postcode={}, 
            %state={},
            country={Brazil}}            
\author{Martin Haardt\fnref{label4} } %% Author name
%\ead{m.haardt@tu-ilmenau.de}
 %\fntext[label4]{}
 %\cortext[cor1]{}

\affiliation[label4]{organization={Communications Research Laboratory (CRL), Technical University Ilmenau},%Department and Organization
            %addressline={}, 
            city={Ilmenau},
            %postcode={}, 
            %state={},
            country={Germany}}

%% Abstract
\begin{abstract}
In this paper, we consider a double-RIS (D-RIS)-aided flat-fading MIMO system and propose an interference-free channel training and estimation protocol, where the two single-reflection links and the one double-reflection link are estimated separately. Specifically, by using the proposed training protocol, the signal measurements of a particular reflection link can be extracted interference-free from \revG{the} measurements of \revG{the superposition of the three links}. We show that some channels are associated with two different components of the received signal. \revgg{Exploiting the common channels involved in the single and double reflection links while recasting the received signals as tensors, we formulate the} coupled tensor-based least square Khatri-Rao factorization (C-KRAFT) algorithm \revG{which is a closed-form solution} and \revG{an enhanced iterative solution with less restrictions on the identifiability constraints, the coupled-alternating least square (C-ALS) algorithm.} The C-KRAFT and C-ALS based channel estimation schemes are used to obtain the channel matrices in both single and double reflection links. \revgg{We show that the proposed coupled tensor decomposition-based channel estimation schemes offer more accurate channel estimates under less restrictive identifiability constraints compared to competing channel estimation methods.} Simulation results are provided showing the effectiveness of the proposed \revgg{algorithms.}
\end{abstract}



%% Keywords

\end{frontmatter}



\section{Introduction}
\label{sec:intro}

Reconfigurable intelligent surfaces (RIS) have been introduced recently as a key enabling technology for future wireless communication systems. An RIS is a 2D surface with a large number of inexpensive tunable elements, e.g., antennas or metamaterials \cite{b1,b2}. Due to these potentials, RIS-aided communication systems have received significant attention in both academia and industry in the last few years. As a result, various RIS-aided communication systems have been studied, such as millimeter-wave (mmWave) communications, dynamic time division duplex (DTDD) based communication systems, cognitive radio, and unmanned aerial vehicle (UAV) communications \cite{b3}--\cite{b7}. 

However, in a passive RIS-aided communication system, where the RIS has no radio-frequency chains, channel estimation is a challenging task since it involves the estimation of multiple channels simultaneously. These include the direct channels between the transmitter and each receiver, the channels
between the RIS and the \revG{transmitter}, and the channels between the RIS and each \revG{receiver} \cite{b8}.  Therefore, most existing \revG{references} about RIS  optimization techniques \cite{b3}--\cite{b9} \revG{assume} that perfect channel state information (CSI) is available at the transceivers. However, this assumption is highly unlikely to hold in practice for an RIS-assisted system. Due to its passive nature, an RIS cannot send and receive pilot symbols and has no signal processing capability to estimate the channels. \revgg{In this case, the channel estimation task should be carried out at the receiver using pilots reflected by the RISs.}

Earlier works on RIS-aided systems have considered single RIS (S-RIS)-aided systems \cite{b10, b11}, where a single RIS is deployed between a user equipment (UE) and a base station (BS). Nonetheless, \revgg{to further exploit the spatial diversity provided by multiple wireless links} multi RIS-aided systems \revG{have} started to receive more attention recently, where two or more RIS \revgg{panels} are deployed between a UE and a BS \cite{b12}--\cite{b14}. In \cite{b15,b16}, the authors showed that double RIS (D-RIS) aided systems have \revG{a} much higher passive beamforming gain when compared to S-RIS aided systems, i.e., $\mathcal{O}(M_{\text{S}}^4)$ versus $\mathcal{O}(M_{\text{S}}^2)$, where $M_{\text{S}}$ denotes the total number of reflective elements in both systems. Channel estimation in D-RIS systems, however, is more complicated than that in S-RIS systems, since the number of channels \revgg{matrices to be estimated} in D-RIS systems is larger \revgg{compared to} S-RIS based systems. 

%On the other hand, in \cite{DRIS_SISO}, two channel estimation methods for D-RIS-aided single-input single-output (SISO) system are proposed considering different inter-RIS channel setups, (i) an arbitrary and (ii) a line-of-sight (LoS) dominant channel. In \cite{DRIS_OnlySingleRefLink}, the linear minimum-mean-square-error (LMMSE) approach is proposed to estimate the channels of both single reflection links and direct link in a single user D-RIS-aided single-input multiple-output (SIMO) system. 

In \cite{b17,b18}, we have proposed  alternating least squares (ALS)-based channel estimation methods for a D-RIS aided MIMO system by \revgg{exploiting two tensor signal structures obtained from} the channel training measurements. However, both methods consider simplified D-RIS aided MIMO systems, where only the double-reflection link from the BS to the UE across the two RISs is available. However, in order to achieve the maximum beamforming gain of a D-RIS-based system, not only the double-reflection link must be estimated, but also the other two single-reflection links. \revG{To this end}, channel estimation methods for D-RIS aided systems considering both single-reflection and double-reflection links have been proposed recently, e.g., \cite{b19}--\cite{b22}. In \cite{b20}, the authors proposed a channel estimation protocol with two-time scales for active D-RIS-aided multi-user multiple-input single-output (MISO) systems, where the \revG{slowly} time-varying channel is estimated by exploiting channel sparsity, and the fast time-varying channel is estimated using a \revG{measurement-augmentation estimate (MAE) compressed sensing strategy. This is based on estimating the fast time-varying channels at the BS using the estimates of the slowly time-varying channels.} 
%In~\cite{DRIS-multiuser_MISO-scaling}, the cascaded channels of both single-reflection and double-reflection links are estimated for a D-RIS-aided multi-user MISO system assuming that the cascaded channel coefficients are scaled version of those of a lower dimensional single-reflection channel. 
In \cite{b21}, a multi-user single-input multiple-output (SIMO) system is considered where the channel estimation procedure is divided into three phases. In the first two phases, one of the two single-reflection links is estimated while turning off the other RIS. In the third phase, the double-reflection link is estimated after canceling the interference from the two single reflection links. 

%

%To the best of our knowledge, there has not been any work on cascaded channel estimation of the both single-reflection links and double-reflection link considering D-RIS-aided MIMO systems.

In this paper, we propose to exploit the benefits of coupled tensor decomposition techniques for channel estimation in a D-RIS-aided MIMO communication system, where the two single-reflection links and the one double-reflection link are estimated \revG{from three different tensors}. \revG{In contrast to} \cite{b23}, our proposed \revG{methods exploit the fact that} some of the channels are common to the double reflection link and the single reflection links. As a result, we exploit these structures and couple the resulting measurement signal \revG{tensors} along the common \revgg{dimension. We derive the decomposition structure of the resulting tensor, from which two new channel estimation algorithms are formulated.} The adopted coupled tensor decomposition technique results in an improvement in the identifiability constraints and enhances the channel estimation accuracy in comparison to \revG{the state-of-the-art} in \cite{b23}. %Specifically, by using the proposed training protocol, we show that the signal measurements of a particular reflection link can be extracted interference-free from measurements of the other reflection links. 

In this way, the MIMO channels involved in a particular reflection link can be estimated  from the extracted measurements using, e.g., the methods \revgg{proposed} in \cite{b10,b18}. \revgg{We show that the interference-free cascaded channel of each single-reflection link} can be represented as a 3-way tensor that \revG{admits a} canonical polyadic (CP) decomposition, while the double-reflection \revgg{cascaded channel} can be represented as a 4-way tensor admitting \revG{a} nested-CP decomposition \cite{ b18,b24}. \revgg{By coupling these received signal tensors, we derive,} coupled tensor-based least square Khatri-Rao factorization (C-KRAFT) and coupled-ALS (C-ALS) based channel estimation schemes to obtain the channel matrices in both single and double reflection links. \revG{In the C-KRAFT scheme, which is a closed-form solution, the two coupled channels are estimated in parallel, while the other two channels are extracted from the coupled terms. Moreover, in the C-ALS scheme, which is an enhanced iterative solution with less restrictions in terms of the identifiability constraints, we estimate one of the channel matrices while assuming that the other channels are fixed.} \revgg{We also derive the identifiability conditions for the proposed channel estimation schemes, showing that the proposed coupled tensor decomposition-based algorithms offer less restrictive training settings compared to competing methods.} Our proposed method is applicable not only to single-antenna UEs, but also to multi-antenna UEs which is different from the work in \cite{b21}. Simulation results are provided showing the effectiveness of the proposed channel training protocol.

\textbf{Notation:} The transpose, the conjugate transpose (Hermitian), the Moore–Penrose inverse, the Kronecker product, and the Khatri-Rao product are denoted as $\bm{X}^{\sf{T}}$, $\bm{X}^{\sf{H}}$, $\bm{X}^{+}$, $\otimes$, $\diamond$, respectively. Moreover, $\text{diag}\{\bm{x}\}$ forms a diagonal matrix $\bm{X}$ by putting the entries of the input vector $\bm{x}$ in its main diagonal, $\text{vec}\{\bm{X}\}$ forms a vector by staking the columns of $\bm{X}$ over each other, $\text{unvec}\{\bm{x}\}$ is the inverse of the vec operator. Furthermore, $\|\bm{a}\|_2$ represents the 2-norm and $\|\bm{A}\|_{\text{F}}$ denotes the Frobenius norm. The expression $\bm{\mathcal{X}}\times_n\bm{A}$ gives the $n$-mode product between a tensor $\bm{\mathcal{X}}\in\mathbb{C}^{I_1\times\cdots I_n\cdots I_R}$ and a matrix $\bm{A}\in\mathbb{C}^{I_n\times J_n}$ that produces the resulting tensor $\bm{\mathcal{Y}}\in\mathbb{C}^{I_1\times\cdots J_n\cdots I_R}$. Additionally, $[\bm{\mathcal{Y}}]_{(n)}$ represents the $n$-mode unfolding of the tensor $\bm{\mathcal{Y}}$.  
\section{System Model}
\label{sec:SysMod}
\begin{figure}
\centering
	\includegraphics[scale = 0.6]{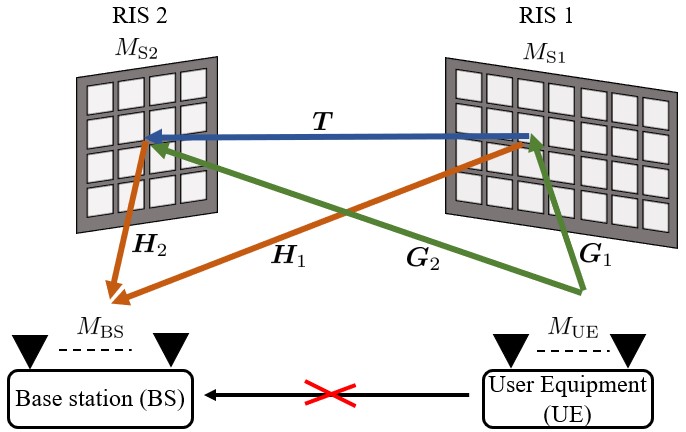}
	\caption{A D-RIS-aided MIMO communication system.}
	\label{fig:SystemModel}
\end{figure}

We consider a D-RIS-aided MIMO communication system as shown in Fig.~\ref{fig:SystemModel}, where \revG{two} RISs, i.e., RIS 1 and RIS 2, are deployed to assist the communication between a UE with $M_{\text{UE}}$ antennas and a BS with $M_{\text{BS}}$ antennas. As depicted in Fig.~\ref{fig:SystemModel}, we assume that RIS~1 with $M_{\text{S}1}$ reflecting elements is placed closer to the BS while RIS~2 with $M_{\text{S}2}$ reflecting elements is placed closer to the UE. Similarly to \cite{b21}, we assume that the direct channel between the BS and the UE is neglectable due to severe signal blockage and pathloss. 

Let $\bm{G}_{1} \in \mathbb{C}^{M_{\text{S}1} \times M_{\text{UE}}}$, $\bm{G}_{2} \in \mathbb{C}^{M_{\text{S}2} \times M_{\text{UE}}}$,  $\bm{H}_{1} \in \mathbb{C}^{M_{\text{BS}} \times M_{\text{S}1}}$, $\bm{H}_{2} \in \mathbb{C}^{M_{\text{BS}} \times M_{\text{S}2}}$, $\bm{T} \in \mathbb{C}^{M_{\text{S}2} \times M_{\text{S}1}}$ denote the UE-to-RIS~1, UE-to-RIS~2, RIS~1-to-BS, RIS~2-to-BS, and RIS~1-to-RIS~2 channels, respectively. Then, the received signal at the $(i,j,k)$th transmission time can be expressed as
\begin{align}
	\bar{\bm{y}}_{i,j,k}   = & \text{ } \bm{H}_{2} \text{diag}\{\bm{\theta}_{2,j}\} \bm{T} \text{diag}\{\bm{\theta}_{1,i}\} \bm{G}_{1} \bm{x}_k  + \bm{H}_1 \text{diag}\{\bm{\theta}_{1,i}\} \bm{G}_{1} \bm{x}_k + \bm{H}_{2} \text{diag}\{\bm{\theta}_{2,j}\} \bm{G}_{2}\bm{x}_k + \bar{\bm{n}}_{i,j,k} \in \mathbb{C}^{M_{\text{BS}}}, \label{eq:MeasSig}
\end{align}
where $\bm{x}_k \in \mathbb{C}^{M_{\text{UE}}}$ is the $k$th training vector at UE with $\|\bm{x}_k\|_2^2 = 1, k \in \{1,\cdots, K\}$, $\bm{\theta}_{1,i} \in \mathbb{C}^{M_{\text{S}1}}$ is the $i$th training beam of RIS~1 with $|[\bm{\theta}_{1,i}]_{[m]}| = \frac{1}{\sqrt{M_{\text{S}1}}}$, $i \in \{1, \cdots, I\}$, $\bm{\theta}_{2,j} \in \mathbb{C}^{M_{\text{S}2}}$ is the $j$th training beam of RIS~2 with $|[\bm{\theta}_{2,j}]_{[n]}| = \frac{1}{\sqrt{M_{\text{S}2}}}, \forall n$, $j \in \{1,\cdots, J\}$, and $\bar{\bm{n}}_{i,j,k} \in \mathbb{C}^{M_{\text{BS}}}$ is the \revG{additive white Guassian noise (AWGN)} with variance $\sigma_n^2$. Note that the received signal at the BS in \eqref{eq:MeasSig} is comprised of three \revG{parts}, \textbf{Component 1} \revG{denotes the single-reflection component} via RIS~1 (second term), \textbf{Component 2} \revG{is the single-reflection component} via RIS~2 (third term), and \textbf{Component 3} \revG{comprises the double-reflection component} via RIS~1 and RIS~2 (the first term). 

Let $\bm{X} = [\bm{x}_1, \cdots, \bm{x}_K] \in \mathbb{C}^{M_{\text{UE}} \times K}$ \revG{denote the training pilot at the transmitter during one frame}. Then, by collecting the received signals $\{\bar{\bm{y}}_{i,j,k}\}_{k = 1}^K$ at the BS next to each other, we obtain 
\begin{align}
\bar{\bm{Y}}_{i,j}  = &  \text{ }\bm{H}_{2} \text{diag}\{\bm{\theta}_{2,j}\} \bm{T} \text{diag}\{\bm{\theta}_{1,i}\} \bm{G}_{1} \bm{X}  + \bm{H}_1 \text{diag}\{\bm{\theta}_{1,i}\} \bm{G}_{1} \bm{X} + \bm{H}_{2} \text{diag}\{\bm{\theta}_{2,j}\} \bm{G}_{2}\bm{X} + \bar{\bm{N}}_{i,j} \in \mathbb{C}^{M_{\text{BS}} \times K}, 
\end{align}
where $\bar{\bm{N}}_{i,j}$ is defined similarly to $\bar{\bm{Y}}_{i,j} $. We assume that $\bm{X}$ is designed with orthonormal rows, i.e., $\bm{X} \bm{X}^{\sf{H}} = \bm{I}_{M_{\text{UE}}}$, which implies that $K \geq M_{\text{UE}}$. Then, the right filtered measurement matrix $\bm{Y}_{i,j} = \bar{\bm{Y}}_{i,j}\bm{X}^{\sf{H}}$ can be expressed as
\begin{align}
\bm{Y}_{i,j} = &  \text{ } \bm{H}_{2} \text{diag}\{\bm{\theta}_{2,j}\} \bm{T} \text{diag}\{\bm{\theta}_{1,i}\} \bm{G}_{1} + \bm{H}_1 \text{diag}\{\bm{\theta}_{1,i}\} \bm{G}_{1} + \bm{H}_{2} \text{diag}\{\bm{\theta}_{2,j}\} \bm{G}_{2} + \bm{N}_{i,j} \in \mathbb{C}^{M_{\text{BS}} \times M_{\text{UE}}}, 
\end{align} 
where $\bm{N}_{i,j}  = \bar{\bm{N}}_{i,j}  \bm{X}^{\sf{H}}$.

\begin{figure}
	\centering
	\includegraphics[scale = 1]{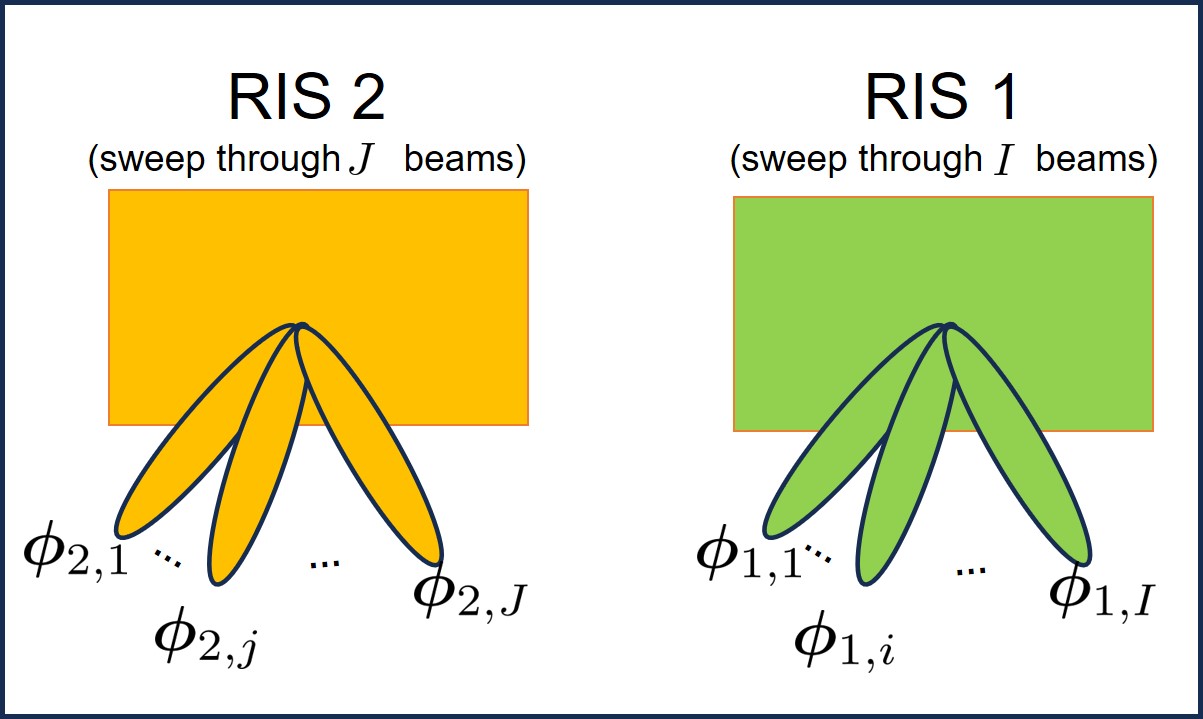}
	\caption{The training procedure.}
	\label{fig:path3}
\end{figure}

%\section{Proposed channel estimation method}
%\label{sec:ChEst}
\revG{The training beams of the RISs} $\bm{\theta}_{1,i}, \forall i$, and $\bm{\theta}_{2,j},\forall j$, can have two states:  State 0 and State 1. In State $0$, the \revG{training} beams are designed as
\begin{equation}
\bm{\theta}^0_{\mu,q} = [e^{j\theta_1},\dots, e^{j\theta_{M_{\text{S}\mu}}}]^{\sf{T}} = \bm{\theta}_{\mu,q} \in \mathbb{C}^{M_{\text{S}\mu}}, \label{eq:State0}
\end{equation}
and in State $1$, the \revG{training} beams are designed as
\begin{equation}
\bm{\theta}^1_{\mu,q} = [e^{j(\theta_1 + \pi)},\dots, e^{j(\theta_{M_{\text{S}\mu}}+ \pi)}]^{\sf{T}}= -\bm{\theta}_{\mu,q} \in \mathbb{C}^{M_{\text{S}\mu}}, \label{eq:State1}
\end{equation} 
with $\mu \in \{1, 2\}$ and $q \in \{i, j\}$. \revG{This training protocol is an extension of the training protocol in \cite{b25} for S-RIS aided systems to the case of D-RIS aided systems.} To estimate the channels, we perform \revG{a} channel training procedure for the RISs as illustrated in Fig.~\ref{fig:path3}. Specifically, for every $i$th training beam of RIS~1, i.e., $\bm{\phi}_{1,i} \in \mathbb{C}^{M_{\text{S}1}}$, $i \in \{1,\dots,I\}$, we sweep through the $J$ training beams of RIS~2, i.e., $\{\bm{\phi}_{2,1}, \dots, \bm{\phi}_{2,J}\}$, where $\bm{\phi}_{2,j} \in \mathbb{C}^{M_{\text{S}2}}, \forall j$. \revG{At this time, the} training beams of both RISs are in State 0. Therefore, the obtained measurement matrices during Stage 1 are given as  
\begin{align}
\bm{Y}_{i,j}^{0,0} & =  \bm{H}_{2} \text{diag}\{\bm{\theta}^{0}_{2,j}\} \bm{T} \text{diag}\{\bm{\theta}^{0}_{1,i}\} \bm{G}_{1}   + \bm{H}_1 \text{diag}\{\bm{\theta}^{0}_{1,i}\} \bm{G}_{1} + \bm{H}_{2} \text{diag}\{\bm{\theta}^{0}_{2,j}\} \bm{G}_{2} + \bm{N}_{i,j}^{0,0}  \in \mathbb{C}^{M_{\text{BS}} \times M_{\text{UE}}}. \label{eq:FiltMeasMat}
\end{align} \revG{Here we use all the combinations of $i,\, \forall i=1,\cdots,I$ and $j,\,\forall j=1,\cdots,J$.}

\begin{figure}
\centering
	\includegraphics[scale = 0.55]{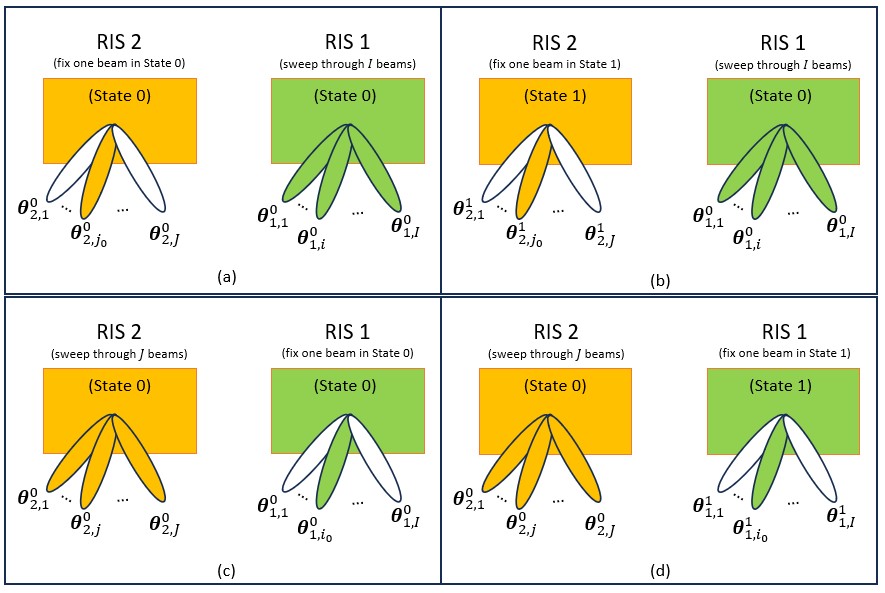}
	\caption{The training procedures to estimate the channels in Component~1 ((a) and (b)) and the channels in  Component~2 ((c) and (d)).}
	\label{fig:path1n2}
\end{figure}

\textbf{Channel training procedure for Component 1:} To estimate  $\bm{G}_{1} $ and  $\bm{H}_{1}$ in Component 1,  as shown in  Fig. \ref{fig:path1n2} (b), we switch the state of a selected training beam of RIS~2 from State 0 to State 1, i.e., $\bm{\theta}_{2,j}^{1}$, while we sweep again through the $I$ training beams of RIS~1, i.e., $\{\bm{\theta}_{1,1}^{0},\dots, \bm{\theta}_{1,I}^{0}\}$. Thus, the obtained measurement matrices during Stage 2 are 
\begin{align} 
	\bm{Y}_{i,j}^{0,1} & = \bm{H}_{2} \text{diag}\{\bm{\theta}^{1}_{2,j}\} \bm{T} \text{diag}\{\bm{\theta}^{0}_{1,i}\} \bm{G}_{1}   + \bm{H}_1 \text{diag}\{\bm{\theta}^{0}_{1,i}\} \bm{G}_{1}  + \bm{H}_{2} \text{diag}\{\bm{\theta}^{1}_{2,j}\} \bm{G}_{2} + \bm{N}_{i,j}^{0,1}  \in \mathbb{C}^{M_{\text{BS}} \times M_{\text{UE}}}  \label{y001}
\end{align}
Therefore, by combining \eqref{eq:FiltMeasMat}  and \eqref{y001} and recalling that $\bm{\theta}_{2,j}^{1} = - \bm{\theta}_{2,j}^{0}$, we obtain the measurement matrices $\bm{Y}_{i,j}^{(1)}$ as 
\begin{equation}
\bm{Y}_{i,j}^{(1)} \triangleq \bm{Y}_{i,j}^{0,0} + \bm{Y}_{i,j}^{0,1} = 2 \bm{H}_1 \text{diag}\{\bm{\theta}^{0}_{1,i}\} \bm{G}_{1} + \bm{N}_{i,j}^{(1)} \label{eq:Get_S1_Step1_Simpl},
\end{equation} 
where $\bm{N}_{i,j}^{(1)} = \bm{N}_{i,j}^{0,0} + \bm{N}_{i,j}^{0,1} $. Note that \eqref{eq:Get_S1_Step1_Simpl} depends only on the training vectors of the RIS~1, i.e., $\{\bm{\theta}_{1,1}^{0},\dots, \bm{\theta}_{1,I}^{0}\}$.

\textbf{Channel training procedure for  Component 2:} To estimate  $\bm{G}_{2} $ and  $\bm{H}_{2}$ in Component 2, we perform another channel training as illustrated  in  Fig. \ref{fig:path1n2} (b), we switch the state of a selected training beam of RIS~1 from State 0 to State~1, i.e., $\bm{\theta}_{1,i}^{1}$, while we sweep again through the $J$ beams of RIS~2, i.e., $\{\bm{\theta}_{2,1}^{0},\dots, \bm{\theta}_{2,J}^{0}\}$ and the obtained measurement matrix is written as 
 \begin{align} 
 \bm{Y}_{i,j}^{1,0} & = \bm{H}_{2} \text{diag}\{\bm{\theta}^{0}_{2,j}\} \bm{T} \text{diag}\{\bm{\theta}^{1}_{1,i}\} \bm{G}_{1}   + \bm{H}_1 \text{diag}\{\bm{\theta}^{1}_{1,i}\} \bm{G}_{1}  + \bm{H}_{2} \text{diag}\{\bm{\theta}^{0}_{2,j}\} \bm{G}_{2} + \bm{N}_{i,j}^{1,0}  \in \mathbb{C}^{M_{\text{BS}} \times M_{\text{UE}}}.  \label{eq:p2b}
\end{align}

Similarly, combining \eqref{eq:FiltMeasMat} and \eqref{eq:p2b}, we obtain the measurement matrix $\bm{Y}_{i,j}^{(2)}$ as 
\begin{align} 
 \bm{Y}_{i,j}^{(2)} \triangleq \bm{Y}_{i,j}^{0,0} + \bm{Y}_{i,j}^{1,0} = 2 \bm{H}_2 \text{diag}\{\bm{\theta}^{0}_{2,j}\} \bm{G}_{2} + \bm{N}_{i,j}^{(2)},  \label{eq:path2}
\end{align} where $\bm{N}_{i,j}^{(2)}=\bm{N}_{i,j}^{0,0} + \bm{N}_{i,j}^{1,0}$. Note that \eqref{eq:path2} is as a result of the fact that $\bm{\theta}_{1,i}^{1} = - \bm{\theta}_{1,i}^{0}$.

% \subsection{Estimate Channel in Component 3}
\textbf{Channel training procedure for Component 3:} To estimate the channel matrix $\bm{T} \in \mathbb{C}^{M_{\text{S}2} \times M_{\text{S}1}}$ in Component 3, we combine the measurement matrices in \eqref{y001} and \eqref{eq:p2b} as follows  
\begin{align}
	{\bm{Y}}_{i,j}^{(3)} & \triangleq -\bm{Y}_{i,j}^{1,0}-\bm{Y}_{i,j}^{0,1}\nonumber\\
 &= 2\bm{H}_{2} \text{diag}\{\bm{\theta}_{2,j}\} \bm{T} \text{diag}\{\bm{\theta}_{1,i}\} \bm{G}_{1} + {\bm{N}}_{i,j}^{(3)}. \label{ydd}
\end{align} \revG{This procedure} will avoid the error propagation in the approach adopted in \cite{b23} for the estimation of this double reflection link.

 The estimates of these channels can be readily obtained by using the methods in \cite{b23}. However, the approach in \cite{b23} has identifiability constraints which in turn affect the estimation accuracy.  Moreover, from Fig.~\ref{fig:SystemModel}, we observe that \revG{the} channels $\bm{G}_1$ and $\bm{H}_2$ are each associated with two different components of the received signal at the BS. Therefore, in the following, we propose to exploit these structures and the inherent \revG{benefits} of coupled tensor \revG{decompositions} to \revG{improve the} identifiability and \revG{the resulting} estimation accuracy. 
\begin{figure}[t]
	\centering
	\includegraphics[scale = 0.80]{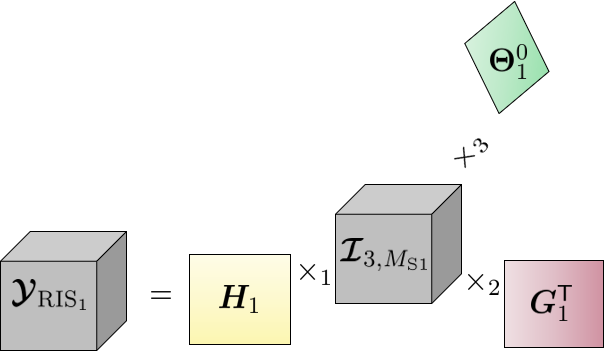}
	\caption{Graphical representation of the 3-way tensor $\bm{\mathcal{Y}}_{\text{RIS}_1}$}
	\label{fig:tens1}
\end{figure}
\section{Tensor Signal Modeling}
In this section, we note that the received measurement matrices for the different components in Section~\ref{sec:SysMod} can be recast as \revG{a} 3-way tensor following CP and nested-CP models. Therefore, by relying on these received signals we are able to jointly estimate all the channels using coupled tensor decomposition algorithms.

\begin{figure}[t]
	\centering
	\includegraphics[scale = 0.80]{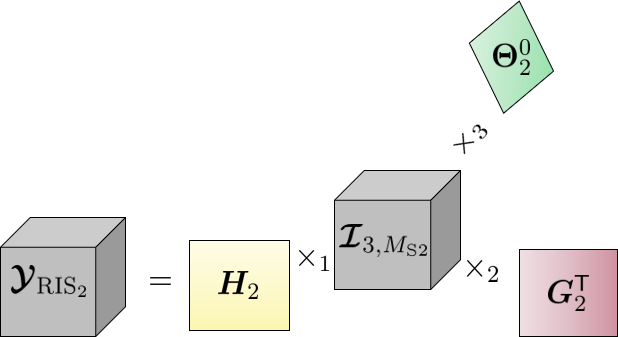}
	\caption{Graphical representation of the 3-way tensor $\bm{\mathcal{Y}}_{\text{RIS}_2}$}
	\label{fig:tens2}
\end{figure}

To begin, we note that specifically 
in \cite{b10}, it is shown that the measurement matrix in \eqref{eq:Get_S1_Step1_Simpl} that only depends on the training beams of RIS~1 can be arranged into a 3-way tensor $\bm{\mathcal{Y}}_{\text{RIS}_1}\in \mathbb{C}^{M_{\text{BS}} \times M_{\text{UE}}\times I}$  admitting a CP decomposition as \begin{align} 
 \bm{\mathcal{Y}}_{\text{RIS}_1} & =\bm{\mathcal{I}}_{3,M_{\text{S1}}} \times_1\bm{H}_{1} \times_2 \bm{G}_{1}^\mathsf{T}\times_3 \bm{\Theta}^0_1+\bm{\mathcal{N}}_{\text{RIS}_1},\label{eq:tens1}  
\end{align} where $\bm{\Theta}^{0}_{1} = [\bm{\theta}^{0}_{1,1}, \cdots, \bm{\theta}^{0}_{1,I}]^{\sf{T}} \in \mathbb{C}^{I \times M_{\text{S}{1}}}$. The tensor $\bm{\mathcal{Y}}_{\text{RIS}_1}$, as shown in Fig.~\ref{fig:tens1}, has the following $n$-mode unfoldings, $n\in\{1,2,3\}$,
\begin{align}
   \big[\bm{\mathcal{Y}}_{\text{RIS}_1}\big]_{(1)}&= \bm{H}_{1} (\bm{\Theta}^0_1\diamond \bm{G}_{1}^\mathsf{T})^\mathsf{T}+\big[\bm{\mathcal{N}}_{\text{RIS}_1}\big]_{(1)}\in \mathbb{C}^{M_{\text{BS}} \times IM_{\text{UE}}}\label{eq:y11}\\
   \big[\bm{\mathcal{Y}}_{\text{RIS}_1}\big]_{(2)}&= \bm{G}_{1}^\mathsf{T} (\bm{\Theta}^0_1\diamond\bm{H}_{1})^\mathsf{T}+\big[\bm{\mathcal{N}}_{\text{RIS}_1}\big]_{(2)}\in \mathbb{C}^{M_{\text{UE}}\times IM_{\text{BS}} }\label{eq:y12}\\
   \big[\bm{\mathcal{Y}}_{\text{RIS}_1}\big]_{(3)}&= \bm{\Theta}^0_1 (\bm{G}_{1}^\mathsf{T}\diamond\bm{H}_{1})^\mathsf{T}+\big[\bm{\mathcal{N}}_{\text{RIS}_1}\big]_{(3)}\in \mathbb{C}^{I\times M_{\text{BS}} M_{\text{UE}}}\label{eq:y13}
\end{align} \revG{In the same way}, we observe that the measurement matrix in \eqref{eq:path2} can be written as a 3-way tensor $\bm{\mathcal{Y}}_{\text{RIS}_2}\in \mathbb{C}^{M_{\text{BS}} \times M_{\text{UE}}\times J}$  admitting a CP decomposition (see Fig.~\ref{fig:tens2})  as
\begin{align} 
 \bm{\mathcal{Y}}_{\text{RIS}_2} & =\bm{\mathcal{I}}_{3,M_{\text{S2}}}\times_1 \bm{H}_{2}\times_2 \bm{G}_{2}^\mathsf{T}  \times_3  \bm{\Theta}^0_2 +\bm{\mathcal{N}}_{\text{RIS}_2},\label{eq:ten2}    
\end{align} where $\bm{\Theta}^{0}_{2} = [\bm{\theta}^{0}_{2,1}, \cdots, \bm{\theta}^{0}_{2,J}]^{\sf{T}} \in \mathbb{C}^{J \times M_{\text{S}{2}}}$ and the $n$-mode unfoldings, $n\in\{1,2,3\}$, can be written as 
\begin{align}
\big[\bm{\mathcal{Y}}_{\text{RIS}_2}\big]_{(1)}&= \bm{H}_{2} (\bm{\Theta}^0_2\diamond\bm{G}_{2}^\mathsf{T})^\mathsf{T}+\big[\bm{\mathcal{N}}_{\text{RIS}_2}\big]_{(1)}\in \mathbb{C}^{M_{\text{BS}} \times JM_{\text{UE}}}\label{eq:y21}\\
   \big[\bm{\mathcal{Y}}_{\text{RIS}_2}\big]_{(2)}&=\bm{G}_{2}^\mathsf{T} (\bm{\Theta}^0_2\diamond\bm{H}_{2})^\mathsf{T}+\big[\bm{\mathcal{N}}_{\text{RIS}_2}\big]_{(2)}\in \mathbb{C}^{M_{\text{UE}}\times JM_{\text{BS}} } \label{eq:y22}\\
   \big[\bm{\mathcal{Y}}_{\text{RIS}_2}\big]_{(3)}&=\bm{\Theta}^0_2 (\bm{G}_{2}^\mathsf{T}\diamond\bm{H}_{2})^\mathsf{T}+\big[\bm{\mathcal{N}}_{\text{RIS}_2}\big]_{(3)}\in \mathbb{C}^{J\times M_{\text{BS}} M_{\text{UE}}} \label{eq:y23}
\end{align}
\begin{figure*}
	\centering
	\includegraphics[scale = 0.6]{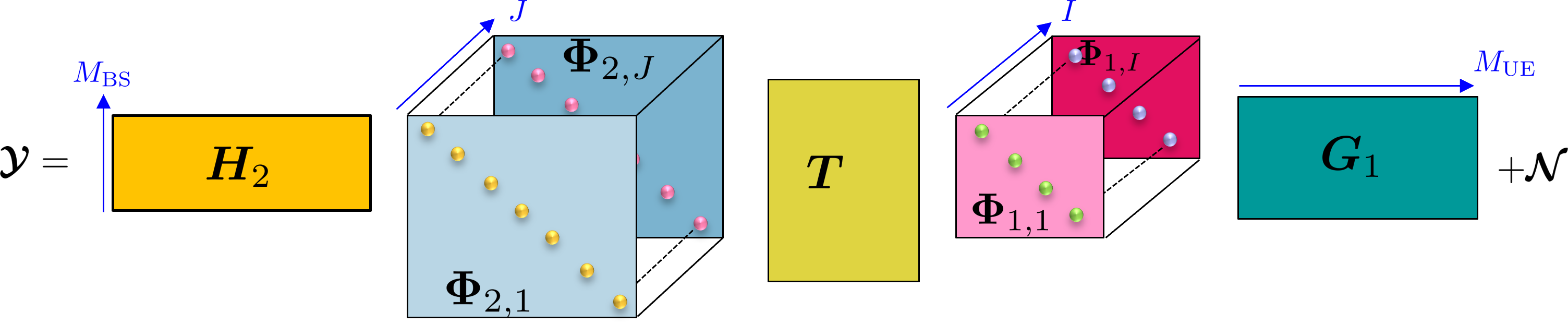}
	\caption{Graphical representation of the 4-way tensor $\bm{\mathcal{Y}}$}
	\label{fig:tens3}
\end{figure*}

Furthermore, we have shown in our previous work \cite{b18} that the measurement matrices ${\bm{Y}}_{i,j}^{(3)}$  in \eqref{ydd} can be arranged into a 3-way tensor $\bm{\mathcal{Y}}$  admitting a nested-CP decomposition as shown in Fig.~\ref{fig:tens3}. Moreover, let $\bm{\mathcal{W}}\in \mathbb{C}^{M_{\text{UE}}\times I \times M_{\text{S}{2}}}$ and $\bm{\mathcal{U}}\in \mathbb{C}^{M_{\text{BS}}\times J \times M_{\text{S}{1}}}$ denote two 3-way CP tensors as 
\begin{align}
    \label{Wtens}\bm{\mathcal{W}}&=\bm{\mathcal{I}}_{3,M_{\text{S1}}} \times_1\bm{G}_{1}^\mathsf{T} \times_2 \bm{\Theta}_1^0\times_3 \bm{T}\in \mathbb{C}^{M_{\text{UE}}\times I \times M_{\text{S}{2}}},\\
    \label{Ztens}\bm{\mathcal{U}}&=\bm{\mathcal{I}}_{3,M_{\text{S2}}} \times_1\bm{H}_{2} \times_2 \bm{\Theta}_2^0\times_3 \bm{T}^\mathsf{T}\in \mathbb{C}^{M_{\text{BS}}\times J \times M_{\text{S}{1}}}.
\end{align}  Therefore, the tensor $\bm{\mathcal{Y}}$ can be written in either of the following two forms depending on the choice of the mode combination of $\bm{\mathcal{Y}}$ with  associated  CP models in \eqref{Wtens} and \eqref{Ztens}, \cite{b26} 
\begin{align*}
    %\label{y11}
    \bm{\mathcal{Y}}^{(1)}&=\bm{\mathcal{I}}_{3,M_{\text{S2}}} \times_1\bm{H}_{2} \times_2\big[\bm{\mathcal{W}}\big]_{(3)}^\mathsf{T}\times_3 \bm{\Theta}_2^0\in \mathbb{C}^{M_{\text{BS}}\times IM_{\text{UE}} \times J}\\
    %\label{y111}
    \bm{\mathcal{Y}}^{(2)}&=\bm{\mathcal{I}}_{3,M_{\text{S1}}} \times_1 \big[\bm{\mathcal{U}}\big]_{(3)}^\mathsf{T}\times_2 \bm{G}_{1}^\mathsf{T}\times_3 \bm{\Theta}_1^0\in \mathbb{C}^{M_{\text{BS}}J\times M_{\text{UE}}\times I },
\end{align*} where $\big[\bm{\mathcal{W}}\big]_{(3)}^\mathsf{T}=\left(\bm{\Theta}_1^0\diamond\bm{G}_{1}^\mathsf{T}\right)\bm{T}^\mathsf{T}\in \mathbb{C}^{IM_{\text{UE}}\times M_{\text{S}2} }$ and $\big[\bm{\mathcal{U}}\big]_{(3)}^\mathsf{T}=\left(\bm{\Theta}_2^0\diamond\bm{H}_{2}\right)\bm{T}\in \mathbb{C}^{JM_{\text{BS}}\times M_{\text{S}1}}$ denotes the transposed 3-mode unfoldings of $\bm{\mathcal{W}}$ and $\bm{\mathcal{U}}$, respectively. 
 Therefore, as shown in \cite{b18} the $n$-mode unfoldings of the tensor $\bm{\mathcal{Y}}^{(1)}$, $n\in\{1,2,3\}$  can be written as 
\begin{align}
    \big[\bm{\mathcal{Y}}^{(1)}\big]_{(1)}&=\bm{H}_{2}\left(\bm{\Theta}_2^0\diamond\big[\bm{\mathcal{W}}\big]_{(3)}^\mathsf{T}\right)^\mathsf{T}+\big[\bm{\mathcal{N}}^{(1)}\big]_{(1)}\in \mathbb{C}^{M_{\text{BS}}\times JIM_{\text{UE}}}\label{eq:ytens1}\\
    \big[\bm{\mathcal{Y}}^{(1)}\big]_{(2)}&=\big[\bm{\mathcal{W}}\big]_{(3)}^\mathsf{T}\left(\bm{\Theta}_2^0\diamond\bm{H}_{2}\right)^\mathsf{T}+\big[\bm{\mathcal{N}}^{(1)}\big]_{(2)}\in \mathbb{C}^{IM_{\text{UE}}\times JM_{\text{BS}}}\\
    \big[\bm{\mathcal{Y}}^{(1)}\big]_{(3)}&=\bm{\Theta}_2^0\left(\big[\bm{\mathcal{W}}\big]_{(3)}^\mathsf{T}\diamond\bm{H}_{2}\right)^\mathsf{T}+\big[\bm{\mathcal{N}}^{(1)}\big]_{(3)}\in \mathbb{C}^{J\times IM_{\text{UE}}M_{\text{BS}}}.
    \end{align} \revG{In the same way}, the $n$-mode unfoldings of \revG{the tensor} $\bm{\mathcal{Y}}^{(2)}$  $n\in\{1,2,3\}$, can be written as
    \begin{align}
\big[\bm{\mathcal{Y}}^{(2)}\big]_{(1)}&=\big[\bm{\mathcal{U}}\big]_{(3)}^\mathsf{T}\left(\bm{\Theta}_1^0\diamond\bm{G}_{1}^\mathsf{T}\right)^\mathsf{T}+\big[\bm{\mathcal{N}}^{(2)}\big]_{(1)}\in \mathbb{C}^{JM_{\text{BS}}\times IM_{\text{UE}}}\\
\big[\bm{\mathcal{Y}}^{(2)}\big]_{(2)}&=\bm{G}_{1}^\mathsf{T}\left(\bm{\Theta}_1^0\diamond\big[\bm{\mathcal{U}}\big]_{(3)}^\mathsf{T}\right)^\mathsf{T}+\big[\bm{\mathcal{N}}^{(2)}\big]_{(2)}\in \mathbb{C}^{M_{\text{UE}}\times IJM_{\text{BS}}}\label{eq:ytens2}\\
\big[\bm{\mathcal{Y}}^{(2)}\big]_{(3)}&=\bm{\Theta}_1^0\left(\bm{G}_{1}^\mathsf{T}\diamond\big[\bm{\mathcal{U}}\big]_{(3)}^\mathsf{T}\right)^\mathsf{T}+\big[\bm{\mathcal{N}}^{(2)}\big]_{(3)}\in \mathbb{C}^{J\times IM_{\text{UE}}M_{\text{BS}}}
    \end{align}
    Here, we observe that there are similarities in the structure of the $2$-mode unfoldings of $\bm{\mathcal{Y}}^{(2)}$ and  $\bm{\mathcal{Y}}_{\text{RIS}_1}$ given in \eqref{eq:y12} and \eqref{eq:ytens2}, respectively, and the $1$-mode unfoldings $\bm{\mathcal{Y}}_{\text{RIS}_2}$ and $\bm{\mathcal{Y}}^{(1)}$ given in \eqref{eq:y21} and \eqref{eq:ytens1}, respectively. In the following, we propose to exploit these structure similarities and adopt coupled tensor decomposition techniques to estimate  $\bm{G}_1$,$\bm{H}_1$, $\bm{G}_2$, $\bm{H}_2$, and $\bm{T}$  in order to improve the identifiability constraints and enhance the channel estimation accuracy.

\subsection{Coupled tensor-based channel estimation}
In this section, relying on the tensor signal models developed in the previous section, we formulate the proposed coupled tensor-based channel estimation algorithm by combining the received signal tensors at the BS, i.e., $\bm{\mathcal{Y}}^{(1)}$, $\bm{\mathcal{Y}}^{(2)}$, $\bm{\mathcal{Y}}_{\text{RIS}_1}$, and $\bm{\mathcal{Y}}_{\text{RIS}_2}$.

\textbf{Estimate channels in Component 1:} From \eqref{eq:y12} and \eqref{eq:ytens2} we discover that the channel matrix $\bm{G}_1$ is common to the 2-mode unfoldings of $\bm{\mathcal{Y}}^{(2)}$ \revG{and} $\bm{\mathcal{Y}}_{\text{RIS}_1}$. Therefore, we denote $\tilde{\bm{Y}}_1$ as a matrix that \revG{stacks} the two 2-mode unfoldings of $\bm{\mathcal{Y}}^{(2)}$ \revG{and} 
 $\bm{\mathcal{Y}}_{\text{RIS}_1}$  along the second dimension, i.e., $\tilde{{\bm{Y}}}_1=\begin{bmatrix}
    \big[\bm{\mathcal{Y}}_{\text{RIS}_1}\big]_{(2)}&\big[\bm{\mathcal{Y}}^{(2)}\big]_{(2)}
\end{bmatrix}\in \mathbb{C}^{M_{\text{UE}}\times I(J+1)M_{\text{BS}}}$ as 
\begin{align}
    \label{eq:ycouple1}
    \tilde{\bm{Y}}_1&=\begin{bmatrix}
        \bm{G}_{1}^\mathsf{T} (\bm{\Theta}^0_1\diamond\bm{H}_{1})^\mathsf{T}&\bm{G}_{1}^\mathsf{T}\left(\bm{\Theta}_1^0\diamond\big[\bm{\mathcal{U}}\big]_{(3)}^\mathsf{T}\right)^\mathsf{T}
    \end{bmatrix},\nonumber\\
    &=\bm{G}_{1}^\mathsf{T}\begin{bmatrix}
        (\bm{\Theta}^0_1\diamond\bm{H}_{1})^\mathsf{T}& \left(\bm{\Theta}_1^0\diamond\big[\bm{\mathcal{U}}\big]_{(3)}^\mathsf{T}\right)^\mathsf{T}
    \end{bmatrix}.
\end{align}We can further express \eqref{eq:ycouple1}  as (see Appendix~A for the proof)
\begin{align}
\bm{Y}_1=\tilde{\bm{Y}}_1{\bm{P}}_1=\bm{G}_{1}^\mathsf{T}\left(\bm{\Theta}_1^0\diamond\bm{\Sigma}_1\right)^\mathsf{T},\label{eq:ycheck}
\end{align}where $\bm{\Sigma}_1=\begin{bmatrix}
        \bm{H}_{1}\\
\big[\bm{\mathcal{U}}\big]_{(3)}^\mathsf{T}
\end{bmatrix}$ and ${\bm{P}}_1$ is a permutation matrix. It should be noted that \eqref{eq:ycheck} can be written as a tensor ${\bm{\mathcal{Y}}}_1\in \mathbb{C}^{M_{\text{UE}}\times (J+1)M_{\text{BS}}\times I}$ which admits a CP decomposition with a coupled nested-CP \revG{model} as 
\begin{align}
    \label{tenscc1}{\bm{\mathcal{Y}}}_1=\bm{\mathcal{I}}_{3,M_{\text{S1}}} \times_1\bm{G}_{1}^\mathsf{T} \times_2 \bm{\Sigma}_{1}\times_3 \bm{\Theta}_1^0+{\bm{\mathcal{N}}}_1.
\end{align} Therefore, we can write the $n$-mode unfoldings of ${\bm{\mathcal{Y}}}_1$, i.e., $n\in\{1,2,3\}$, as 
\begin{align} \Big[{\bm{\mathcal{Y}}}_1\Big]_{(1)}&=\bm{G}_{1}^\mathsf{T}\left(\bm{\Theta}_1^0\diamond\bm{\Sigma}_1\right)^\mathsf{T}+\Big[{\bm{\mathcal{N}}}_1\Big]_{(1)}\in \mathbb{C}^{M_{\text{UE}}\times I(J+1)M_{\text{BS}}}\label{y1a}\\
\Big[{\bm{\mathcal{Y}}}_1\Big]_{(2)}&=\bm{\Sigma}_1\left(\bm{\Theta}_1^0\diamond\bm{G}_{1}^\mathsf{T}\right)^\mathsf{T}+\Big[{\bm{\mathcal{N}}}_1\Big]_{(2)}\in \mathbb{C}^{(J+1)M_{\text{BS}}\times IM_{\text{UE}}}\label{y1b}\\
\Big[{\bm{\mathcal{Y}}}_1\Big]_{(3)}&=\bm{\Theta}_1^0\left(\bm{\Sigma}_1\diamond\bm{G}_{1}^\mathsf{T}\right)^\mathsf{T}+\Big[{\bm{\mathcal{N}}}_1\Big]_{(3)}\in \mathbb{C}^{I\times (J+1)M_{\text{BS}}M_{\text{UE}}}\label{y1c}.
\end{align}

 \textbf{Estimate channels in Component 2:} To proceed, we note that the channel matrix $\bm{H}_2$ is common to the 1-mode unfoldings of $\bm{\mathcal{Y}}^{(1)}$ \revG{and} $\bm{\mathcal{Y}}_{\text{RIS}_2}$ given in \eqref{eq:ytens1} and \eqref{eq:y21}, respectively. Similarly, we define $\tilde{\bm{Y}}_2$ as a matrix that \revG{stacks} the 1-mode unfoldings of $\bm{\mathcal{Y}}$ \revG{and} $\bm{\mathcal{Y}}_{\text{RIS}_2}$ along the second dimension, i.e., $\tilde{\bm{Y}}_2=\begin{bmatrix}
    \big[\bm{\mathcal{Y}}_{\text{RIS}_2}\big]_{(1)}&\big[\bm{\mathcal{Y}}^{(1)}\big]_{(1)}
\end{bmatrix}\in \mathbb{C}^{M_{\text{BS}}\times J(I+1)M_{\text{UE}}}$ as 
\begin{align}\label{eq:ycheck2}
    \tilde{\bm{Y}}_2&=\begin{bmatrix}
        \bm{H}_{2} (\bm{\Theta} ^0_2\diamond\bm{G}_{2}^\mathsf{T})^\mathsf{T}&\bm{H}_{2}\left(\bm{\Theta}_2^0\diamond\big[\bm{\mathcal{W}}\big]_{(3)}^\mathsf{T}\right)^\mathsf{T}
    \end{bmatrix},\nonumber\\
    &=\bm{H}_{2}\begin{bmatrix} (\bm{\Theta}^0_2\diamond\bm{G}_{2}^\mathsf{T})^\mathsf{T}&\left(\bm{\Theta}_2^0\diamond\big[\bm{\mathcal{W}}\big]_{(3)}^\mathsf{T}\right)^\mathsf{T}
    \end{bmatrix}.
\end{align} Then, we can further express \eqref{eq:ycheck2} in a similar manner as \eqref{eq:ycheck} as 
\begin{align}
    \label{eq:ycheck22}\bm{Y}_2=\tilde{\bm{Y}}_2{\bm{P}}_2=\bm{H}_{2}\left(\bm{\Theta}_2^0\diamond\bm{\Sigma}_2\right)^\mathsf{T},
\end{align}where $\bm{\Sigma}_2=\begin{bmatrix}
            \bm{G}_{2}^\mathsf{T}\\
\big[\bm{\mathcal{W}}\big]_{(3)}^\mathsf{T}
        \end{bmatrix} $ and ${\bm{P}}_2$ is a permutation matrix. The measurement matrix in \eqref{eq:ycheck22} can also be written as a tensor ${\bm{\mathcal{Y}}}_2\in \mathbb{C}^{M_{\text{BS}}\times (I+1)M_{\text{UE}} \times  J}$ which admits a CP decomposition with a coupled nested-CP \revG{structure} as \begin{align}
    \label{tenscc2}{\bm{\mathcal{Y}}}_2=\bm{\mathcal{I}}_{3,M_{\text{S2}}} \times_1\bm{H}_{2} \times_2 \bm{\Sigma}_{2}\times_3 \bm{\Theta}_2^0+{\bm{\mathcal{N}}}_2,
\end{align} and its $n$-mode unfoldings, i.e., $n\in\{1,2,3\}$, are given as  \begin{align}
    \Big[{\bm{\mathcal{Y}}}_2\Big]_{(1)}&=\bm{H}_{2}\left(\bm{\Theta}_2^0\diamond\bm{\Sigma}_2\right)^\mathsf{T}+\Big[{\bm{\mathcal{N}}}_2\Big]_{(1)}\in \mathbb{C}^{M_{\text{BS}}\times J(I+1)M_{\text{UE}}}\label{y2a}\\
    \Big[{\bm{\mathcal{Y}}}_2\Big]_{(2)}&=\bm{\Sigma}_2\left(\bm{\Theta}_2^0\diamond\bm{H}_{2}\right)^\mathsf{T}+\Big[{\bm{\mathcal{N}}}_2\Big]_{(2)}\in \mathbb{C}^{(I+1)M_{\text{UE}}\times JM_{\text{BS}}}\label{y2b}\\
    \Big[{\bm{\mathcal{Y}}}_2\Big]_{(3)}&=\bm{\Theta}_2^0\left(\bm{\Sigma}_2\diamond\bm{H}_{2}\right)^\mathsf{T}+\Big[{\bm{\mathcal{N}}}_2\Big]_{(3)}\in \mathbb{C}^{J\times (I+1)M_{\text{UE}}JM_{\text{BS}}}\label{y2c}
\end{align}

        \textbf{Estimate channel in Component 3:} The channel $\bm{T}\in \mathbb{C}^{M_{\text{S}2}\times M_{\text{S}1}}$ in \textbf{Component 3} can easily be estimated by adopting the approach in \cite{b24}. We note that the measurement matrix in \eqref{ydd} is a generalized unfolding of a 4-way tensor admitting a nested-CP decomposition, which can be utilized to estimate the channel matrix $\bm{T}$, i.e., $\bm{Y}_{i,j}^{(3)}=\Big[\bm{\mathcal{Y}}_3\Big]_{([1,2],[3,4])}$. Therefore, the generalized unfolding \cite{b27} is given as  
\begin{align}\label{eq:t}
    \Big[\bm{\mathcal{Y}}_3\Big]_{([1,2],[3,4])} 
&=(\bm{\Theta}_2^0 \diamond {\bm{H}}_2) \bm{T} (\bm{\Theta}_1^0 \diamond {\bm{G}}_1^{\sf{T}})^{\sf{T}}  \in\mathbb{C}^{JM_\text{BS} \times I M_\text{UE} }, 
\end{align}where the each factor $(\bm{\Theta}_2^0 \diamond {\bm{H}}_2)$ and $(\bm{\Theta}_1^0 \diamond {\bm{G}}_1^{\sf{T}})$ results from the combination of the first two modes of the $\bm{\mathcal{U}}$ and $\bm{\mathcal{W}}$ tensor, respectively, whereas the middle factor $\bm{T}$ is the common factor to the CP decompositions of these tensors. %\revG{We can write the measurement matrix in \eqref{eq:t} as a 4-way nested CP tensor $\bm{\mathcal{Y}}_3\in\mathbb{C}^{M_{\text{BS}} \times J \times M_{\text{UE}} \times I }$ as} 
%\begin{align}
   % \bm{\mathcal{Y}}_3={\bm{\mathcal{T}}}\times_1{\bm{H}}_2\times_2\bm{\Theta}_2^0\times_3{\bm{G}}_1^{\sf{T}}\times_4\bm{\Theta}_1^0+\bm{\mathcal{N}}_3
%\end{align}where the core tensor ${\bm{\mathcal{T}}}$  is a super diagonal tensor that is given as ${\bm{\mathcal{T}}}=\text{diag}\{\text{vec}\{\bm{T}\}\}$. 
In the following, we propose two solutions for estimating $\bm{G}_1$,$\bm{H}_1$, $\bm{G}_2$, $\bm{H}_2$, and $\bm{T}$ by exploiting the above tensor structures. %The channels can be estimated as a solution to the following problem
%\begin{align}
    %\hat{\bm{G}}_1,\hat{\bm{H}}_1,\hat{\bm{G}}_2,\hat{\bm{H}}_2,\hat{\bm{T}}=\underset{(\bm{G}_1,\bm{H}_1,\bm{G}_2,\bm{H}_2,\bm{T})}{\text{argmin}} J_{\text{tot}},
%\end{align} where $J_{\text{tot}}$ is the objective function which is given as \begin{align*}
    %J_{\text{tot}}&=\left\Vert\bm{\mathcal{Y}}_{1}-\llbracket  \bm{G}_1^{\sf{T}},\bm{\Sigma}_1,\bm{\Theta}_1^0\rrbracket\right\Vert_{\text{F}}^2+\left\Vert\bm{\mathcal{Y}}_{2}-\llbracket \bm{H}_2 ,\bm{\Sigma}_2,\bm{\Theta}_2^0\rrbracket\right\Vert_{\text{F}}^2+\left\Vert\bm{\mathcal{Y}}_{3}-\Tilde{\bm{\mathcal{T}}}\times_1{\bm{H}}_2\times_2{\bm{G}}_1^{\sf{T}}\times_3\bm{\Theta}_1^0\times_4\bm{\Theta}_2^0\right\Vert_{\text{F}}^2,
%\end{align*} \revG{where $\llbracket\bm{A},\bm{B},\bm{C}\rrbracket$ denotes a  CP tensor with $\bm{A},\bm{B}$, and $\bm{C}$ as the factor matrices, i.e., \begin{align*}
    %\llbracket\bm{A},\bm{B},\bm{C}\rrbracket\triangleq\bm{\mathcal{I}}_M\times_1\bm{A}\times_2\bm{B}\times_3\bm{C},
%\end{align*} where $\bm{A}\in\mathbb{C}^{P\times M}$, $\bm{B}\in\mathbb{C}^{Q\times M}$, and $\bm{C}\in\mathbb{C}^{R\times M}$.}

\subsubsection{Coupled-Khatri-Rao Factorization (C-KRAFT) }
Assume that the training matrices $\bm{\Theta}_1^0$ and $\bm{\Theta}_2^0$ are designed with orthonormal columns, i.e., $(\bm{\Theta}_1^0)^\mathsf{H}\bm{\Theta}_1^0{=}\bm{I}_I$ and $(\bm{\Theta}_2^0)^\mathsf{H}\bm{\Theta}_2^0=\bm{I}_J$. Then, the left-filtered 3-mode unfoldings of ${\bm{\mathcal{Y}}}_1$  in \eqref{y1c} and ${\bm{\mathcal{Y}}}_2$ in \eqref{y2c} are given 
as
\begin{align}
    \label{ybar1}\Big[\bar{\bm{\mathcal{Y}}}_1\Big]_{(3)}&=(\bm{\Theta}_1^0)^\mathsf{H}\Big[{\bm{\mathcal{Y}}}_1\Big]_{(3)}=(\bm{\Sigma}_1\diamond{\bm{G}}_1^{\sf{T}})^{\sf{T}}+\Big[\bar{\bm{\mathcal{N}}}_1\Big]_{(3)}\\
    \label{ybar2}\Big[\bar{\bm{\mathcal{Y}}}_2\Big]_{(3)}&=(\bm{\Theta}_2^0)^\mathsf{H}\Big[{\bm{\mathcal{Y}}}_2\Big]_{(3)}=(\bm{\Sigma}_2\diamond{\bm{H}}_2)^{\sf{T}}+\Big[\bar{\bm{\mathcal{N}}}_2\Big]_{(3)},
\end{align} \revG{where $\Big[\bar{\bm{\mathcal{N}}}_1\Big]_{(3)}{=}(\bm{\Theta}_1^0)^\mathsf{H}\Big[{\bm{\mathcal{N}}}_1\Big]_{(3)}$ and $\Big[\bar{\bm{\mathcal{N}}}_2\Big]_{(3)}{=}(\bm{\Theta}_2^0)^\mathsf{H}\Big[{\bm{\mathcal{N}}}_2\Big]_{(3)}$ are the left-filtered noise matrices. }

Therefore, the estimates of $\bm{G}_1$, $\bm{\Sigma}_1$, $\bm{H}_2$, and $\bm{\Sigma}_2$ can be obtained as a solution to following problems, using $\Big[\bar{\bm{\mathcal{Y}}}_1\Big]_{(3)}$ and $\Big[\bar{\bm{\mathcal{Y}}}_2\Big]_{(3)}$, respectively, \begin{align}\label{hatcomp1}
    (\hat{\bm{G}}_1,\hat{\bm{\Sigma}}_1)&=\underset{\bm{G}_1,\bm{\Sigma}_1}{\text{argmin}}\left\Vert\Big[\bar{\bm{\mathcal{Y}}}_1\Big]_{(3)}^{\sf{T}}-(\bm{\Sigma}_1\diamond{\bm{G}}_1^{\sf{T}})\right\Vert_{\text{F}}^2\\
    (\hat{\bm{H}}_2,\hat{\bm{\Sigma}}_2)&=\underset{\bm{H}_2,\bm{\Sigma}_2}{\text{argmin}}\left\Vert\Big[\bar{\bm{\mathcal{Y}}}_2\Big]_{(3)}^{\sf{T}}-(\bm{\Sigma}_2\diamond{\bm{H}}_2)\right\Vert_{\text{F}}^2.\label{hatcomp2}
\end{align}Closed-form solutions to the above problems are easily obtained by using the least squares Khatrio-Rao Factorization (KRF) technique \cite{b28}--\cite{b30}. Then, $\hat{\bm{H}}_1$ and $\hat{\bm{G}}_2$ can be obtained  from $\hat{\bm{\Sigma}}_1$ and $\hat{\bm{\Sigma}}_2$ by extracting the first $M_{\text{BS}}$ rows of $\hat{\bm{\Sigma}}_1$ and first $M_{\text{UE}}$ rows of $\hat{\bm{\Sigma}}_2$,   respectively as,
$\hat{\bm{H}}_1=\Big[\hat{\bm{\Sigma}}_1\Big]_{[1:M_{\text{BS}},:]}$ and $\hat{\bm{G}}_2=\Big[\hat{\bm{\Sigma}}_2\Big]_{[1:M_{\text{UE}},:]}$. Then, to obtain the estimate of $\bm{T}$ we apply bilinear filtering to \eqref{eq:t}  as 
\begin{align}\label{compo3}
    \hat{\bm{T}}=\Big[(\bm{\Theta}_1^0 \diamond \hat{\bm{G}}_1^{\sf{T}})\Big]^+\Big[\bm{\mathcal{Y}}_3\Big]_{([1,2],[3,4])} ^{\sf{T}}\Big[(\bm{\Theta}_2^0 \diamond \hat{\bm{H}}_2)^{\sf{T}}\Big]^+.
\end{align} 
\begin{figure}[t]
	\centering
	\includegraphics[scale = 0.55]{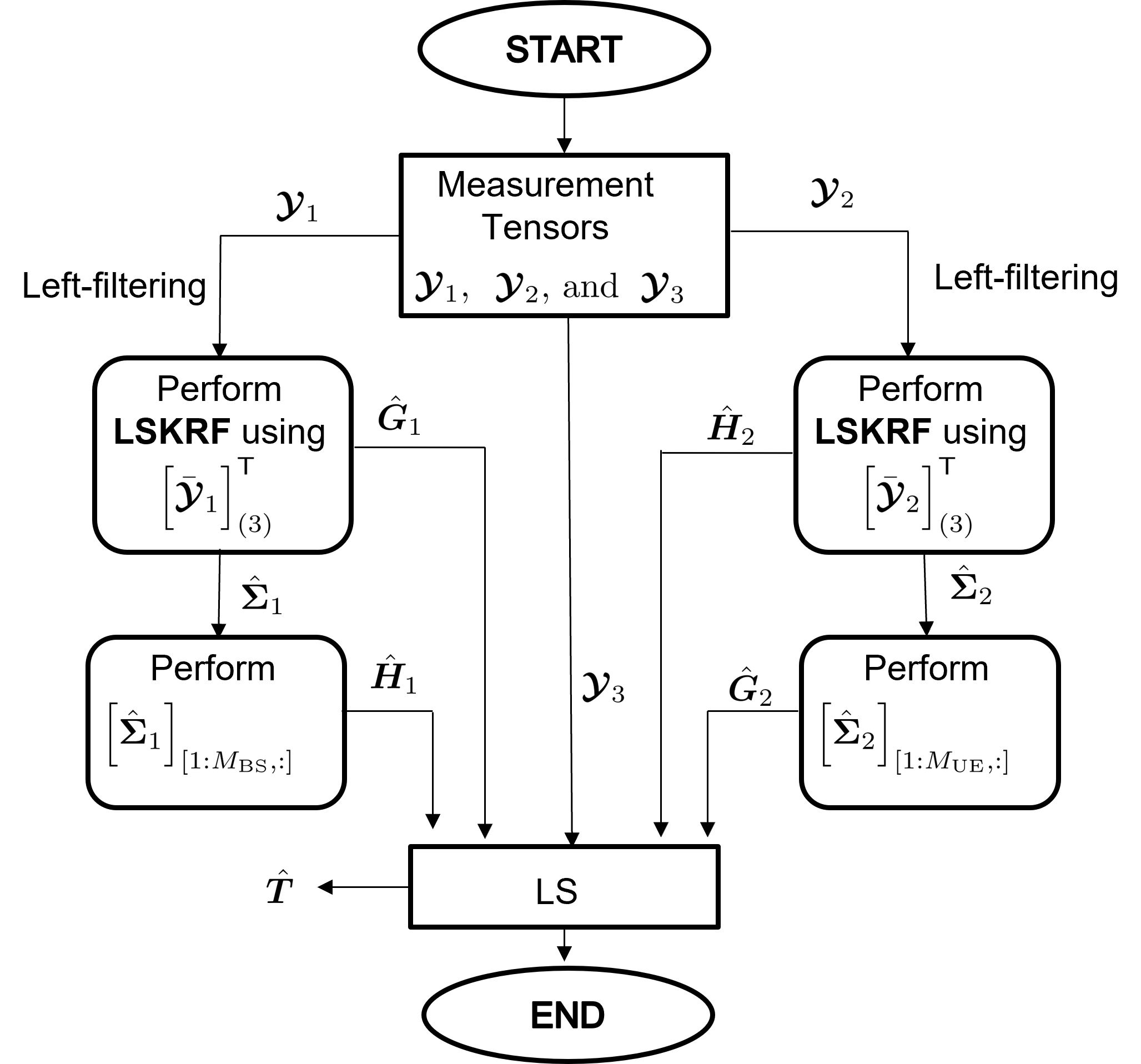}
	\caption{Graphical representation of the C-KRAFT algorithm}
	\label{fig:ckraft}
\end{figure}
The proposed closed-form C-KRAFT-based channel estimation for double-RIS aided MIMO systems is summarized in Fig.~\ref{fig:ckraft} and  Algorithm~\ref{CKRAFT}. The C-KRAFT-based method can be computationally less expensive, however, it may not give a better estimation result. Motivated by the need to improve the accuracy of the channel estimation, in the following, we propose a coupled tensor-based ALS method for the estimation of the channels.
\begin{algorithm}[!htb]
 	\caption{C-KRAFT-based CE for DRIS Systems}
	\label{CKRAFT}
	\begin{algorithmic}[1]
		\State{Input: Measurement tensors $\bm{\mathcal{Y}}_{1} $, $\bm{\mathcal{Y}}_{2}$}
		\State{Get $\hat{\bm{G}}_1$ and $\hat{\bm{\Sigma}}_1$ as a solution to \eqref{hatcomp1} using \cite[Algorithm 1]{b29}}
  \State{Get $\hat{\bm{H}}_2$ and $\hat{\bm{\Sigma}}_2$ as a solution to \eqref{hatcomp2} using \cite[Algorithm 1]{b29}}
  \State{Get $\hat{\bm{H}}_1$ as
  $\hat{\bm{H}}_1=\Big[\hat{\bm{\Sigma}}_1\Big]_{[1:M_{\text{BS}},:]}$}
  \State{Get $\hat{\bm{G}}_2$ as 
  $\hat{\bm{G}}_2=\Big[\hat{\bm{\Sigma}}_2\Big]_{[1:M_{\text{UE}},:]}$ }
  \State{Get $\hat{\bm{T}}$ as in \eqref{compo3}
   using the given $\hat{\bm{G}}_1$ and $\hat{\bm{H}}_2$}
		\State{ Output: $\hat{\bm{G}}_{1}$, $\hat{\bm{H}}_{1}$,$\hat{\bm{H}}_{2}$, $\hat{\bm{G}}_{2}$, and $\hat{\bm{T}}$}
	\end{algorithmic}
\end{algorithm} 

\subsubsection{Coupled-Alternating Least Squares (C-ALS)}
In this section, to improve the estimation accuracy we propose a coupled tensor-based ALS (C-ALS) algorithm which minimizes the data fitting error with respect to one of the channels, with the other channels being fixed. To begin, we note that from the 1-mode \revG{unfolding} of the coupled tensor ${\bm{\mathcal{Y}}}_1$ in \eqref{tenscc1} the common channel  $\bm{G}_1$  can be estimated  using least square (LS) as
\begin{align}
    \hat{\bm{G}}_1=\Big\{\Big[{\bm{\mathcal{Y}}}_1\Big]_{(1)}\begin{bmatrix}
        \left(\bm{\Theta}_1^0\diamond\bm{\Sigma}_{1}\right)^\mathsf{T}
    \end{bmatrix}^+\Big\}^\mathsf{T}.\label{eq:g1hat}
\end{align}
 Then, the estimate of the remaining channel in component 1 is obtained by exploiting \eqref{eq:y11} since the channel is not coupled to any other links. \revG{The LS estimate of} ${\bm{H}}_1$ is obtained as \begin{align}\label{eq:h1hat}
     \hat{\bm{H}}_1 = \Big[\bm{\mathcal{Y}}_{\text{RIS}_{1}}\Big]_{(1)}\Big[ \left(\bm{\Theta}^{0}_{1} \diamond \bm{G}_{1}^{\sf{T}} \right)^{\sf{T}} \Big]^{+}. 
 \end{align}

Next, we consider the estimate of the other common channel $\bm{H}_2$ by exploiting the 1-mode \revG{unfolding} of the coupled tensor ${\bm{\mathcal{Y}}}_2$ in \eqref{y2a}.  As a result, \revG{the LS estimate} of $\bm{H}_2$ can be obtained  as \begin{align}
            \label{eq:h2hat}\hat{\bm{H}}_2=\Big[{\bm{\mathcal{Y}}}_2\Big]_{(1)}\begin{bmatrix}
                \left(\bm{\Sigma}_{2}\diamond\bm{\Theta}_2^0\right)^\mathsf{T}
            \end{bmatrix}^+.
        \end{align}\revG{The LS estimate} of $\bm{G}_2$ is obtained from \eqref{eq:y22} as \begin{align}\label{eq:g2hat}
            \hat{\bm{G}}_2=\Big\{\big[\bm{\mathcal{Y}}_{\text{RIS}_2}\big]_{(2)} \Big[\left(\bm{\Theta}^0_2\diamond\bm{H}_{2}\right)^\mathsf{T}\Big]^+\Big\}^\mathsf{T}.
        \end{align}

\revG{Finally,} we update $\hat{\bm{T}}$ by using \eqref{compo3}. Therefore, the coupled tensor-based ALS channel estimation for a double-RIS aided MIMO system is summarized in Algorithm~\ref{ALS_SingleReflection_DRIS}, which is guaranteed to converge monotonically to a local optimum point \cite{b31}. 
\begin{algorithm}[t]
 	\caption{C-ALS-based CE for DRIS Systems }
	\label{ALS_SingleReflection_DRIS}
	\begin{algorithmic}[1]
		\State{Input: Measurement tensors $\bm{\mathcal{Y}}_{\text{RIS}_{1}} $, $\bm{\mathcal{Y}}_{\text{RIS}_{2}}$, 
 $\bm{\mathcal{Y}}$, and select $t_{\max}$ }
		\State{Initialize: $\hat{\bm{H}}^{(0)}_{1}$, $\hat{\bm{H}}^{(0)}_{2}$, $\hat{\bm{G}}^{(0)}_{2}$, $\hat{\bm{T}}^{(0)}$, e.g., as solution of C-KRAFT algorithm . }
		\For{$t = 1$ to $t_{\max}$}
		\State{$\hat{\bm{G}}_1^{(t)}=\Big\{\Big[{\bm{\mathcal{Y}}}_1\Big]_{(1)}\begin{bmatrix}
        \left(\bm{\Theta}_1^0\diamond\bm{\Sigma}_{1}^{(t-1)}\right)^\mathsf{T}
    \end{bmatrix}^+\Big\}^\mathsf{T}$}
		\State{$\hat{\bm{H}}_1^{(t)} = \Big[\bm{\mathcal{Y}}_{\text{RIS}_{1}}\Big]_{(1)}\Big[ \left( \bm{\Theta}^{0}_{1} \diamond (\hat{\bm{G}}_{1}^{(t)})^{\sf{T}} \right)^{\sf{T}} \Big]^{+}$}
  \State{$\hat{\bm{H}}_2^{(t)}=\Big[{\bm{\mathcal{Y}}}_2\Big]_{(1)}\begin{bmatrix}
                \left(\bm{\Theta}_2^0\diamond\bm{\Sigma}_{2}^{(t-1)}\right)^\mathsf{T}
            \end{bmatrix}^+$}
            \State{$ \hat{\bm{G}}_2^{(t)}=\Big\{\big[\bm{\mathcal{Y}}_{\text{RIS}_2}\big]_{(2)} \Big[\left(\bm{\Theta}^0_2\diamond\bm{H}_{2}^{(t)}\right)^\mathsf{T}\Big]^+\Big\}^\mathsf{T}$}
            \State{$\hat{\bm{T}}^{(t)}=\Big[(\bm{\Theta}_1^0 \diamond \hat{(\bm{G}}_1^{(t)})^{\sf{T}})\Big]^+\Big[\bm{\mathcal{Y}}_3\Big]_{([1,2],[3,4])} ^{\sf{T}}\Big[(\bm{\Theta}_2^0 \diamond \hat{\bm{H}}_2^{(t)})^{\sf{T}}\Big]^+$}
		\EndFor	
		\State{ Output: $\hat{\bm{G}}_{1}$, $\hat{\bm{H}}_{1}$,$\hat{\bm{H}}_{2}$, $\hat{\bm{G}}_{2}$, and $\hat{\bm{T}}$}
	\end{algorithmic}
\end{algorithm} 
\subsection{Identifiablity conditions}
As far as the estimation of the single reflection links and the double reflection link are concerned, the study of identifiability conditions of the associated tensor models is relevant. Indeed, these conditions indicate the required system setups that lead to a unique estimation of the involved channel matrices. To this end, in the following, we derive a set of conditions involving system parameters such as the required number of training frames, and the number of receive and transmit antennas for accurate channel estimation. 
\subsubsection{C-KRAFT based method}\label{3.2.1}
In this section, we \revG{provide} the identifiability conditions associated with the estimation of the D-RIS channels by using the C-KRAFT method in Algorithm~\ref{CKRAFT}. The C-KRAFT based algorithm in LS sense requires that $K \geq M_{\text{UE}}$, $I\geq M_{\text{S}1}$ and $J\geq M_{\text{S}2}$  for accurate estimation of $\bm{G}_1$ and $\bm{\Sigma}_1$ and $\bm{G}_2$ and $\bm{\Sigma}_2$, respectively.

\subsubsection{C-ALS based method}
In this section, we give the identifiablity conditions associated with the estimation of the D-RIS channels by using the C-ALS method in  Algorithm~\ref{ALS_SingleReflection_DRIS}.

\textbf{Component 1:} \revG{The LS estimation of $\bm{G}_1$ and $\bm{H}_1$} requires that each of the following matrices  $(\bm{\Theta}_1^0\diamond\bm{\Sigma}_{1})^\mathsf{T}\in \mathbb{C}^{M_{\text{S}1}\times I(J+1)M_{\text{BS}}}$ in \eqref{eq:g1hat} and $\big( \bm{\Theta}^{0}_{1} \diamond \bm{G}_{1}^{\sf{T}} \big)^{\sf{T}}\in \mathbb{C}^{IM_{\text{UE}}\times M_{\text{S}1}} $ in \eqref{eq:h1hat} \revG{has} a unique right Moore-Penrose pseudo-inverse, i.e., full row-rank. This implies that  $I$ needs to satisfy the condition of $I \geq \text{max} \left\{ \lceil \frac{M_{\text{S}1}}{M_{\text{UE}}} \rceil, \lceil \frac{M_{\text{S}1}}{(J+1)M_{\text{BS}}} \rceil \right\}$ and $K \geq M_{\text{UE}}$ to have accurate channel estimates in the LS sense. This demonstrates a significant gain of $(J+1)$ in the identifiability condition by using this coupled tensor approach \revG{as compared to the state-of-the-art in \cite{b23}.}   

\textbf{Component 2:} 
By investigating \eqref{eq:h2hat} and \eqref{eq:g2hat}, we note that to have accurate channel estimates \revG{of $\bm{G}_2$ and $\bm{H}_2$} in LS sense  $J$ needs to satisfy the condition of  $J \geq \text{max} \left\{ \lceil \frac{M_{\text{S}2}}{(I+1)M_{\text{UE}}} \rceil, \lceil \frac{M_{\text{S}2}}{M_{\text{BS}}} \rceil \right\}$ and $K \geq M_{\text{UE}}$ to have accurate channel estimates in the LS sense. \revG{In the same way}, due to the coupled tensor approach adopted in this work, we obtain an improvement in the identifiability condition for \revG{the} channels in components 2 in comparison to the work in \cite{b23} where $J$ needs to satisfy the condition   $J \geq \text{max} \left\{ \lceil \frac{M_{\text{S}2}}{M_{\text{UE}}} \rceil, \lceil \frac{M_{\text{S}2}}{M_{\text{BS}}} \rceil,2\right\}$ to have accurate channel estimates in the LS sense.

\textbf{Component 3:}
Note that the training procedure shown in Fig. \ref{fig:path3} implies that $IJK$ training overhead is required when estimating \revG{the} channel $\bm{T}$ in Component 3. By investigating \eqref{compo3}, $I$ and $J$ need to satisfy the conditions  $I \geq \lceil \frac{M_{\text{S}1}}{M_{\text{UE}}} \rceil$ and $J \geq \lceil \frac{M_{\text{S}2}}{M_{\text{BS}}} \rceil$, respectively, to have an accurate channel estimate, assuming that $\text{rank}\{ \bm{\Theta}_{1}^0\} = I$ and $\text{rank}\{ \bm{\Theta}_{2}^0\} = J$.

We have summarized the identifiability conditions and the computational cost of the considered algorithms in Table~\ref{tab:sample_table}, where $t_{\text{max}}$ is the number of maximum iterations for the ALS-based schemes. We note that from the measurement tensors in \eqref{tenscc1} and \eqref{tenscc2}, both $I$ and $J$  need to be greater or equal to 2, i.e., $I\geq 2$ and $J\geq 2$. 
\begin{table*}[t]
\centering
\caption{Identifiablity and Computational Cost}
\begin{tabular}{|p{2cm}|p{5cm}|p{5cm}|}
\hline
Method & Identifiablity & Computational Cost \\ \hline
C-KRAFT & $K \geq M_{\text{UE}}$, $I\geq M_{\text{S}1}$, $J\geq M_{\text{S}2}$  & $\mathcal{O}\left(2M_{\text{S}1}^2+2M_{\text{S}2}^2\right)$ \\ \hline
C-ALS & $I \geq \text{max} \{ \lceil \frac{M_{\text{S}1}}{M_{\text{UE}}} \rceil, \lceil \frac{M_{\text{S}1}}{(J+1)M_{\text{BS}}} \rceil \}$, $J \geq \text{max} \{ \lceil \frac{M_{\text{S}2}}{(I+1)M_{\text{UE}}} \rceil, \lceil \frac{M_{\text{S}2}}{M_{\text{BS}}} \rceil \}$, $K \geq M_{\text{UE}}$ & $\mathcal{O}\left(t_{\text{max}}\Big(2M_{\text{S}1}^3+2M_{\text{S}2}^3\Big)\right)$ \\ \hline
ALS \cite{b23} & $J \geq \text{max} \{ \lceil \frac{M_{\text{S}2}}{M_{\text{UE}}} \rceil, \lceil \frac{M_{\text{S}2}}{M_{\text{BS}}} \rceil ,2\}$, $I \geq \text{max} \{ \lceil \frac{M_{\text{S}1}}{M_{\text{UE}}} \rceil, \lceil \frac{M_{\text{S}1}}{M_{\text{BS}}} \rceil ,2\}$,  $K \geq M_{\text{UE}}$ & $\mathcal{O}\left(t_{\text{max}}\Big(2M_{\text{S}1}^3+2M_{\text{S}2}^3+(M_{\text{S}1}M_{\text{S}2})^3\Big)\right)$  \\ \hline
\end{tabular}
\label{tab:sample_table}
\end{table*}

\textbf{Ambiguities:}
The ambiguities of the estimated channels in each of the components can be determined using a similar approach \revG{as} in \cite{b17}. It was shown in \cite{b17} that the estimated channels are unique up to scalar ambiguities per column. For example,  these ambiguities between the estimated and the true channels in component 3 can be written as

$$
\hat{\boldsymbol{T}} \approx \boldsymbol{\Delta}_{\mathrm{2}}^{-1} \boldsymbol{T} \boldsymbol{\Delta}_{\mathrm{1}}^{-1}, \quad \hat{\boldsymbol{H}}_{2} \approx \boldsymbol{H}_{2} \boldsymbol{\Delta}_{2}, \quad \hat{\boldsymbol{G}}_{1} \approx \boldsymbol{\Delta}_{\mathrm{1}} \boldsymbol{G}_{\mathrm{1}}
$$
where $\boldsymbol{\Delta}_{\mathrm{1}}$ and $\boldsymbol{\Delta}_{\mathrm{2}}$ are diagonal matrices holding the scaling ambiguities. However, these ambiguities have no impact on the reconstructed estimate of the effective end-to-end channel $\bm{H}_2\bm{T}\bm{G}_1$ as their effects disappear. Note that, due to the knowledge of the RIS reflection matrices $\bm{\Theta}_1^0$ and $\bm{\Theta}_2^0$ at the receiver, the permutation ambiguities do not exist \cite{b10}. The ambiguities in the estimated channels in the other components can be determined by following a similar approach.  %Moreover,the method in \cite{nwalozieCO} requires $3IJK$ training overhead to estimate the channels which can be too high for large networks. However, due to the new training sequences for the RISs given in \eqref{newRIS1} and \eqref{newRIS2} the channel training procedure can be carried out only once, and the resulting  measurement matrix in \eqref{eq:FiltMeasMat} will have all the different combinations. As a result, this requires $IJK$ training overhead for both the C-KRAFT and C-ALS algorithms.
\begin{figure}[!htb]
	\centering
	\begin{subfigure}[b]{0.45\textwidth}
		\centering
		\includegraphics[width=\linewidth]{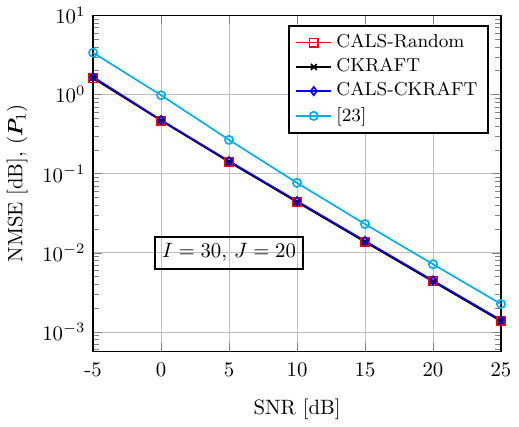}  
		\caption[network2]%
		 {{\small NMSE vs SNR for $\bm{P}_1$}}
		\label{fig:c1}
	\end{subfigure}
\hfil
	\begin{subfigure}[b]{0.45\textwidth}
		\includegraphics[width=\linewidth]{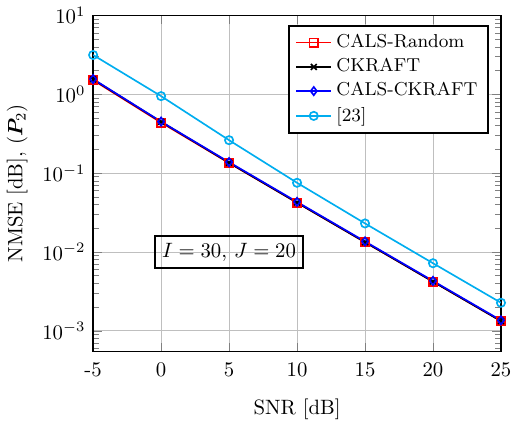}  
		\caption[]%
		{{\small NMSE vs SNR for $\bm{P}_2$}}
		\label{fig:c2}
	\end{subfigure}
	%\hfil
	\begin{subfigure}[b]{0.45\textwidth}
    \centering
		\includegraphics[width=\linewidth]{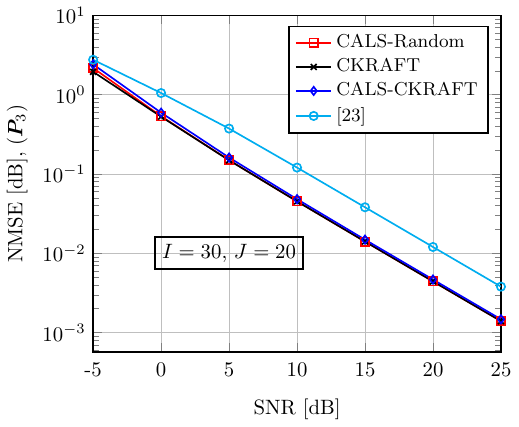} 
		\caption[]%
		{{\small NMSE vs SNR for $\bm{P}_3$}}
		\label{fig:c3}
	\end{subfigure}
    %\hspace{5cm}
	\caption[]
	{\small NMSE vs SNR of the cascaded channels, $M_{\text{S}1}=30,\,M_{\text{S}2}=20,\,\{I,J\}=\{30,20\}$}
	\label{fig:main}
\end{figure}

\section{Simulation Results}
\label{sec:SimulationResults}

We assume that the entries of $\bm{G}_{1}$, $\bm{G}_{2}$, $\bm{H}_{1}$, $\bm{H}_{2}$, and $\bm{T}$ are independent and identically distributed zero-mean circularly-symmetric complex Gaussian random variables. We show results in terms of the normalized-mean-squared-error (NMSE) of the cascaded channels as 
\begin{align}
	\text{NMSE} =  \frac{\mathbb{E} \{ \|\hat{\bm{P}}_{x} - {\bm{P}}_{x}\|_{\text{F}}^2\}}{\mathbb{E} \{\|{\bm{P}}_{x}\|_{\text{F}}^2\}},
\end{align}
where $\bm{P}_x$ is the effective cascaded channel for component $x$, and  $x \in \{1,2,3\}$. Moreover, $\bm{P}_\text{1} = \bm{H}_1 \bm{G}_1$, $\bm{P}_\text{2} = \bm{H}_2 \bm{G}_2$, and $\bm{P}_\text{3} = \bm{H}_2 \bm{T}\bm{G}_1$. We design the training matrices at the RISs, i.e., $\bm{\Theta}^{0}_1$, and $\bm{\Theta}^{0}_2$ as truncated DFT matrices.   \revG{In all the simulations,} we assume that $M_{\text{BS}} = 4$, $M_{\text{UE}} = 2$, $K=2$, $M_{\text{S}1} = 30$, and $M_{\text{S}2} = 20$.
\begin{figure}[!htb]
		\centering
	\begin{subfigure}[b]{0.45\textwidth}
		\centering
		\includegraphics[width=\linewidth]{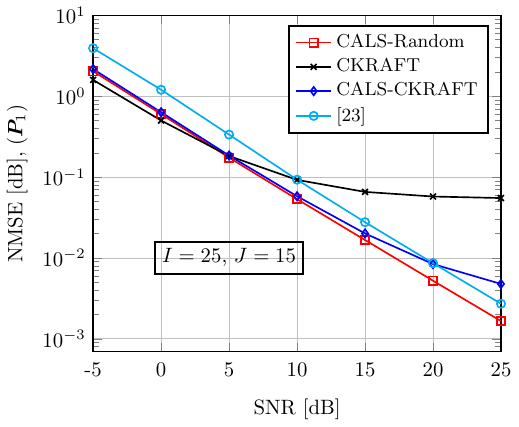}  
		\caption[network2]%
		 {{\small NMSE vs SNR for $\bm{P}_1$}}
		\label{fig:cc1}
	\end{subfigure}
\hfil
	\begin{subfigure}[b]{0.45\textwidth}
		\includegraphics[width=\linewidth]{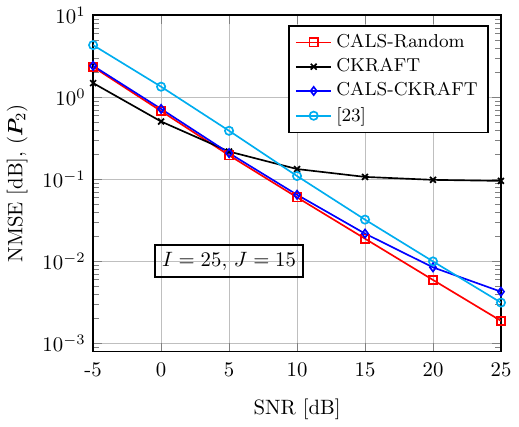}  
		\caption[]%
		{{\small NMSE vs SNR for $\bm{P}_2$}}
		\label{fig:cc2}
	\end{subfigure}
	%\hfil
	\begin{subfigure}[b]{0.45\textwidth}
    \centering
		\includegraphics[width=\linewidth]{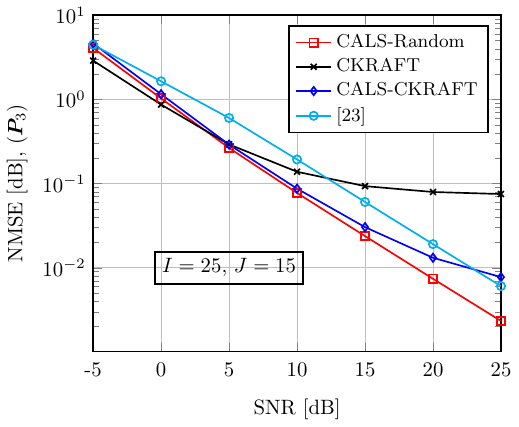} 
		\caption[]%
		{{\small NMSE vs SNR for $\bm{P}_3$}}
		\label{fig:cc3}
	\end{subfigure}
    %\hspace{5cm}
	\caption[]
	{\small NMSE vs SNR of the cascaded channels, $M_{\text{S}1}=30,\,M_{\text{S}2}=20,\,\{I,J\}=\{25,15\}$}
	\label{fig:main2}
\end{figure}

In Figures~\ref{fig:c1}~-~\ref{fig:c3}, we show \revG{the} NMSE versus \revG{the} signal-to-noise ratio (SNR) for the cascaded \revG{channels} in components 1, 2, and 3, respectively, comparing the two proposed channel estimation methods, the CKRAFT method, i.e., Algorithm~\ref{CKRAFT}, the C-ALS method, i.e., Algorithm~\ref{ALS_SingleReflection_DRIS}, and the channel estimation method in \cite{b23}.  From the figures, we can see clearly that when $I$ and $J$ are selected such that $I=M_{\text{S}1}$ and $J=M_{\text{S}2}$, the proposed coupled tensor-based channel estimation frameworks, i.e., CKRAFT, Algorithm~\ref{ALS_SingleReflection_DRIS} with CKRAFT initialization and Algorithm~\ref{ALS_SingleReflection_DRIS} with random initialization have a better performance in comparison with the scheme in \cite{b23}. This performance improvement is as a result of the adopted coupled tensor decompositions which brought about significant gain in terms of the identifiablity constraints. \revG{The results show that C-KRAFT and C-ALS with random initialization have a similar performance. This implies that we do not have to use the C-ALS algorithm if the identifiability conditions for the C-KRAFT algorithm are satisfied. This is due to the fact that the C-ALS algorithm is computationally more expensive.}

In Figures~\ref{fig:cc1}~-~\ref{fig:cc3}, we show \revG{the} NMSE versus \revG{the} SNR  for the cascaded \revG{channels} in components 1, 2, and 3, respectively, while selecting $I< M_{\text{S}1}$ and $J< M_{\text{S}2}$. From the figures, we see that the performance of the CKRAFT algorithm degrades significantly. This performance degradation is due to the fact that the identifiability constraints for the CKRAFT algorithm given in Section~\ref{3.2.1} are not satisfied. However, the CALS algorithm with random initialization which is less restrictive in terms of the identifiability condition achieves a good performance at \revG{the} expense of a higher complexity in terms of \revG{the} number of iterations.  The CALS algorithm attains \revG{a} good performance only after a few iterations, i.e., $t_{\text{max}}=10$.

\section{Conclusions}
\label{sec:conclusion}
In this paper, we \revG{consider} the channel estimation problem in D-RIS-aided flat-fading MIMO systems. To reduce the signaling overhead, we \revG{propose} an interference-free channel training protocol, which allows us to extract the signal measurements of a particular reflection link interference-free from measurements of \revG{the superposition of the three links}. To further improve the identifiability constraints and improve \revG{the} channel estimation accuracy, we exploit the structure of the common channels and the inherent benefits of coupled tensor decompositions. \revG{We propose a closed-form solution based on the least square Khatrio-Rao factorization and an enhanced iterative solution using an alternating least square approach.   Using the coupled tensor-based least square Khatri-Rao factorization (C-KRAFT) and the coupled-ALS (C-ALS) based channel estimation schemes, the channel matrices in both single and double reflection links are obtained. The C-ALS algorithm is less restrictive in terms of the identifiability condition at the expense of additional complexity with respect to the number of iterations. The results show that C-KRAFT and C-ALS with random initialization have a similar performance. This implies that we do not have to use the C-ALS algorithm if the identifiability conditions for the C-KRAFT algorithm are satisfied.  The provided simulation results show the effectiveness of the proposed channel training protocol.}

\appendix
\section{DETAILED DERIVATION OF \eqref{eq:ycheck}}
\label{app1}
To begin, we note the following identities, Property 1: $\text{vec}\{\bm{X} \bm{Y} \bm{Z}\} = (\bm{Z}^{\sf{T}} \otimes \bm{X}) \text{vec}\{\bm{Y}\}$. Property 2: $\text{vec}\{\bm{X} \text{diag}\{\bm{y}\} \bm{Z}\} = (\bm{Z}^{\sf{T}} \diamond \bm{X}) \bm{y}$. Property~3: $\begin{bmatrix}
    \bm{A}\diamond\bm{C}\\
    \bm{A}\diamond\bm{B}
\end{bmatrix}=\bm{P}\left(\bm{A}\diamond\begin{bmatrix}
    \bm{C}\\
    \bm{B}
\end{bmatrix}\right)$, where $\bm{P}$ is a permutation matrix. Property~4: $\begin{bmatrix}
    \bm{A}\\
    \bm{B}
\end{bmatrix}^\mathsf{T}=\begin{bmatrix}
    \bm{A}^\mathsf{T}&\bm{B}^\mathsf{T}\
\end{bmatrix}$. Then, the coupled measurement matrix in \eqref{eq:ycouple1} which is given as 
\begin{align}
    %\label{eq:ycouple1}
    \tilde{\bm{Y}}_1&=\bm{G}_{1}^\mathsf{T}\begin{bmatrix}
        \Big(\bm{Q}_1\diamond\bm{\Theta}^0_1\diamond\bm{H}_{1}\Big)^\mathsf{T}& \left(\bm{\Theta}_1^0\diamond\big[\bm{\mathcal{U}}\big]_{(3)}^\mathsf{T}\right)^\mathsf{T}
    \end{bmatrix},
\end{align} can be expressed as 
\begin{align}
    \tilde{\bm{Y}}_1&=\begin{bmatrix}
        \Big\{\Big(\bm{\Theta}^0_1\diamond\bm{H}_{1}\Big)\bm{G}_{1}\Big\}^\mathsf{T}&\Big\{\left(\bm{\Theta}_1^0\diamond\big[\bm{\mathcal{U}}\big]_{(3)}^\mathsf{T}\right)\bm{G}_{1}\Big\}^\mathsf{T}
    \end{bmatrix}\overset{(a)}{=}\begin{bmatrix}
        \Big\{\Big(\bm{\Theta}^0_1\diamond\bm{H}_{1}\Big)\bm{G}_{1}\Big\}\\
        \Big\{\left(\bm{\Theta}_1^0\diamond\big[\bm{\mathcal{U}}\big]_{(3)}^\mathsf{T}\right)\bm{G}_{1}\Big\}
    \end{bmatrix}^\mathsf{T}\nonumber\\
    &\overset{(b)}{=}\begin{bmatrix}
        \bm{P}\begin{bmatrix}\bm{\Theta}_1^0\diamond\underbrace{
            \begin{bmatrix}
                \bm{H}_{1}\\
                \big[\bm{\mathcal{U}}\big]_{(3)}^\mathsf{T}     \end{bmatrix}}_{\bm{\Sigma}_1}\end{bmatrix}\bm{G}_{1}
    \end{bmatrix}^\mathsf{T}=\Big\{{\bm{P}}\Big[\bm{\Theta}_1^0\diamond\bm{\Sigma}_1\Big]\bm{G}_{1}\Big\}^\mathsf{T}=\bm{G}_{1}^\mathsf{T}\Big[\bm{\Theta}_1^0\diamond\bm{\Sigma}_1\Big]^\mathsf{T}{\bm{P}}^\mathsf{T},
\end{align}where $(a)$ is due to the application of Property~4, $(b)$ is obtained by using Property~4, and  ${\bm{P}}$ is a permutation matrix.

%\bibliographystyle{elsarticle-num} 
%\bibliography{refs}



\end{document}